\documentclass[a4paper,10pt]{article}
\usepackage[utf8]{inputenc}

\title{Emergence of Fractional Kinetics in Spiny Dendrites}
\author{Silvia Vitali,  Francesco Mainardi and Gastone Castellani}

\begin{document}

\maketitle

\begin{abstract}
Fractional extensions of the cable equation have been proposed in the literature to describe transmembrane potential in spiny dendrites. 
The anomalous behavior has been related in the literature
to the geometrical properties of the system, in particular, the density of spines, by experiments, computer
simulations, and in comb-like models.~The same PDE can be related to more than one stochastic process leading to anomalous diffusion behavior. 
The time-fractional diffusion equation can be associated to a continuous time random walk (CTRW) with power-law waiting time probability or to a special case of the Erdély-Kober
fractional diffusion, described by the ggBm. 
In this work, we show that time fractional generalization of 
the cable equation arises naturally in the CTRW by considering a superposition of Markovian processes and in a {\it ggBm-like} 
 construction of the random variable.
\end{abstract}

\section{Introduction}

Neurons are the fundamental structural units of the nervous system. These cells are specialized to communicate with each other through electrical and chemical signals,
specifically called neural signals. 
Despite the incredible diversity existing between different neuron types, the basic mechanism to exchange electrical signals is the same 
as for other excitable cells, and is driven by transmembrane ion currents, generating a variation in the transmembrane voltage $V_m$.
The cell membrane is composed by a phospholipid bilayer that isolates the inner part of the cell from the surround, crossed by several macromolecular structures (proteins) that allow 
ions and other molecules to flow in and out. These~biological constituents have been historically described by the use of electrical circuit elements to model
the transmembrane potential of the cells.
A capacitor element was introduced to mimic the role of the phospholipid bilayer, which keeps a different charge concentration in and out the cell.
To~model the presence of embedded proteins was considered a resistance, in parallel to the capacitor, to roughly describe the flow of charges in and out. 

Dendrites and axons are nonisopotentential parts of the neuron in which the membrane can be geometrically approximated by a cylinder
, with the longitudinal axis length greater than the radius. Neglecting radial flow phenomena, 
the cell membrane is described by a linear density of $RC$  modules, 
 composed by a transmembrane capacitance $c_m$ and resistance $r_m$ in parallel, 
connected by an~internal (and~eventually external) resistance $r_i$ associated to the ionic flow parallel to the membrane in the cytoplasm viscous medium 
inside the cell (or the extracellular fluid outside).~{The differential equation describing the voltage in this circuit system 
is a diffusion equation with the addition of an extra term that accounts for
the decay of the electric signal in time:}
\begin{equation}\label{linear-scaled}
  \frac{\partial V_m(X,T)}{\partial T}= \frac{\partial^2V_m(X,T)}{\partial X^2}-V_m(X,T)\,,
\end{equation}

where, following \cite{Magin}, the adimensional variables $X=x/\lambda$ and $T=t/\tau$ are introduced. The constants $\lambda=\sqrt{r_m/r_i}$ and $\tau=r_m c_m$, the space and time scales of the process, are
determined by the values of the membrane resistance and capacitance per unit length of the system.
The resulting fundamental solution of the Chauchy problem is a Gaussian suppressed by an exponential decay:
\begin{equation}\label{clascab}
 V_{m}(X,T)=\frac{1}{\sqrt{4\pi T}}e^{-(\frac{X^2}{4T} + T)}\,,
\end{equation}
the appearance of exponential decay is determined by the term $-V_m$ in the equation, 
which accounts in fact for particles lost in the system through the transmembrane channels. 
 In this system, the mean square displacement (MSD) of diffusing molecules is expected to scale linearly with time.
The motion of ions and the subsequent propagation of the action potential along the axons is well described by this model; however, 
anomalous diffusion behavior in the propagation of the 
subthreshold potential in dendrites has been measured by several experiments \cite{Nimchinsky-2002,Ionescu,Jacobs-1997}.
 Experimental evidences of anomalous diffusion of an inert tracer in spiny branches of Purkinje cells \cite{Fide-2006} suggested that the origin
of anomalous diffusion in this system was related to the geometry of the system, and that anomalous diffusion was related to the presence of spines
more than to the presence of branches. Furthermore, spine density can change dynamically depending on neuronal activity. 
The idea of correlation between spines density and the anomalous time scaling exponent of the MSD was suggested
 in \cite{Henry2008}, because spine density is an~important feature for the physiological behavior of several types of neurons, and the 
low density of spines is associated to aging \cite{Jacobs-1997,Duan-2003},  neurological disorders \cite{Nimchinsky-2002} and syndromes  \cite{Motohir-1980}.

In \cite{Henry2008}, the anomalous diffusion of ions was introduced modifying the Nernst-Planck equation (NPE)
to generate a fractional Brownian motion (fBm) and a continuous time random walk (CTRW) process.
In these cases the diffusion coefficient $D$ is not constant as in the standard NPE, but it becomes a time dependent operator {characterized by }
the scaling parameter 
$0<\alpha \leq1$. 
For fBm 
 and CTRW respectively we have:
\begin{equation}
 D(\alpha,t)_{fBm}= D(\alpha)_{fBm}\alpha t^{\alpha-1}\,,\quad
 D(\alpha,t)_{CTRW}= D(\alpha)_{CTRW}\frac{\partial^{1-\alpha}}{\partial t^{1-\alpha}}\,,
\end{equation}
where $\frac{\partial^{1-\alpha}}{\partial t^{1-\alpha}}$ is the Riemann--Liouville fractional derivative operator. 

This approach leads to the following differential equations for the two models:
\begin{equation}\label{cablefbm}
 \frac{\partial V_{fBm}}{\partial T}=\alpha T^{\alpha -1} \frac{\partial^2 V_{fBm}}{\partial X^2} -\mu^2\kappa T^{\kappa -1}V_{fBm}\,,
\end{equation}
and
\begin{equation}
 \frac{\partial V_{CTRW}}{\partial T}=\frac{\partial^{\alpha-1}}{\partial T^{\alpha -1}} \frac{\partial^2 V_{CTRW}}{\partial X^2}
 -\mu^2\frac{\partial^{\kappa -1}V_{CTRW}}{\partial T^{\kappa -1}}\,,
\end{equation}
where the terms $-\mu^2\kappa T^{\kappa -1}V_{fBm}$ and $ -\mu^2\frac{\partial^{\kappa -1}V_{CTRW}}{\partial T^{\kappa -1}}$ still account for particles loss in the system
instead of the term $-V_m(X,T)$ in Equation (\ref{linear-scaled}).
The exponential decay in time associated to the fundamental solution is still evident in the case of fBm model:
\begin{equation}\label{fbm}
V_{fBm}(X,T)= \frac{1}{\sqrt{4\pi T^\alpha}}e^{-\left(\frac{X^2}{4 T^\alpha}+\mu^2 T^\kappa\right)}\,,
\end{equation}
while it is difficult to notice it for the CTRW model.

Studying the behavior of the fundamental solutions for $V_m$ for both the fBm and CTRW models in~\cite{Henry2008} 
it was determined that subdiffusive behavior enlarges the window of high potential at the soma, despite it lowers the maximum value of the peak, { with
respect to the classic solution in Equation (\ref{clascab}),
and it was suggested \cite{Henry2008} that this effect could be in fact helpful to counterbalance deferred postsynaptic potentials over dendrites and to reduce
temporal attenuation of the signal.}
Then, high spine density should have been related to more enhanced subdiffusive behavior.

More recently, other experiments have been performed on 
 Purkinje cells and pyramidal cells~\cite{Santamaria-2011} and the correlation between spine density and anomalous diffusion exponent in these types of neurons
was explicitly studied; the anomalous MSD was described in terms of an exponent $d_\omega$ by the introduction
of a time dependent diffusion coefficient:
\begin{equation}
 \langle X^2(t)\rangle = 2D(t)\times t = \Gamma\times t ^{2/d_w}\,,
\end{equation}
and linear correlation was found between the parameter $d_\omega$ and the measured density of spines in both the types of neurons studied.
{The effect of the system geometry on the transport regime in dendrites has been modeled considering the geometrical similarities between a comb
structure and a spiny dendrite by the application} of comb-like models of diffusion \cite{Iomin-2013,Mendez-2013}. 
In this model, it was considered that particles may diffuse in both spines, the fingers of the comb, and the dendrite, the backbone of the comb,
where~spines behave as a traps for the moving particles, 
and the average survival time $\tau$ inside each spine is determined by its geometry.
Markovian process was assumed inside each spine, i.e., exponential distribution of survival time
$\Psi_M(t,\tau)=\int_t^\infty\frac{1}{\tau}e^{-t'/\tau}dt'=e^{-t/\tau}$, but the random size and shape of the
spines \cite{Nimchinsky-2002} entail that the final process is the sum of many independent Markovian 
processes averaged over the distribution of the timescale $\tau$ \cite{Iomin-2013,Mendez-2013}:
\begin{equation}
 \Psi(t)=\int_0^\infty \Psi_M(t,\tau)f_S(\tau)d\tau \,.
\end{equation}

When $f_S(\tau)$ is a power law $\Psi(t)$ shows a power law behavior as well, and subdiffusive diffusion appears.
{In the present paper, we develop two models, CTRW and generalized grey Brownian motion (ggBm)
 like, to model diffusion in
the system under study. Since the experiments described in the literature were performed using inert
fluorescent tracers, the models proposed were kept as simple as possible and based on a diffusion
process with leakage, to account for the loss of particles from the system. Subdiffusive behavior was
included in both the models by means of the heterogeneity of the environment.
}
\section{Results}
The emergence of fractional kinetics in complex media in CTRW
was introduced more explicitly as a general concept in \cite{Pagnini-2014}. Analogously to the comb-like model presented, in that short note
the special case of a survival probability of the Mittag-Leffer type was there derived
in terms of a Markovian process with characteristic waiting time properly distributed:
\begin{equation}
\int_0^\infty \Psi_M(t,\tau)f_S(\tau)d\tau =E_\alpha(-t^{\alpha})\,,
\end{equation}
where $E_\alpha(\cdot)$ is the one parameter Mittag-Leffer function \cite{MainardiE}:
\begin{equation}
 E_\alpha(z):=\sum_{n=0}^{\infty}
 \frac{ z^n}{ \Gamma[\alpha  n + 1 ]}\,,\quad
 \alpha>0\,,
\end{equation}

and $f_S(\tau)=\frac{1}{\tau^2}K_\alpha(1/\tau)$. {The distribution $K_\alpha=K_{\alpha,\alpha}^{-\alpha}$ is the fundamental solution 
of the space-time fractional diffusion equation \cite{mainardi_etal-fcaa-2001} for the special case in which fractional orders of derivation of the space and time variables 
are equal, and of order $\alpha$, with extremal asymmetry parameter equal to $-\alpha$, leading to a solution defined on the positive real axes only.}
Within this approach $f_S(\tau)$ corresponds to the stationary distribution of these timescales.
If instead we consider the non stationary case, it holds for the general case:
\begin{equation}
 \Psi(t)=\int_0^\infty \Psi_M(t,\tau)f(\tau,t)d\tau\,,
\end{equation}
however, in the non-stationary case the solution for $f(\tau,t)$ could be no unique given $\Psi(t)$. 

In the present case, the following identity holds:
\begin{equation}
\Psi(t) = \int_0^\infty \Psi_M(t,\tau)f(\tau,t)d\tau=\int_0^\infty e^{-q}\mathcal{H}(q,t)dq\,,
\end{equation}
 then for $\Psi(t)=E_\alpha(-t^{\alpha})$ we may write:
\begin{equation}\label{f-q}
 \mathcal{H}(q,t)=\frac{1}{t^\alpha}M_\alpha(q/t^\alpha)\,,
\end{equation}
or equivalently:
\begin{equation}
 f(\tau,t)=\frac{1}{\tau^2}t^{1-\alpha}M_\alpha(t^{1-\alpha}/\tau)\,,
\end{equation}
where $M_\alpha(z)=W_{-\alpha, 1-\alpha}(z)$ is the $M$-Wright 
 function, special case of the Wright function $ W_{\lambda, \mu }(z)$ defined by the series \cite{MainardiF}:
\begin{equation} \label{Wright-function}
 W_{\lambda, \mu }(z) := \sum_{n=0}^{\infty}
 \frac{ z^n}{  n!\, \Gamma[\lambda  n + \mu ]}\,,\quad
 \lambda>-1, \; \mu\ge 0\,.
 \end{equation}
 
 The relation in Equation (\ref{f-q}) is a consequence of the Laplace transform relation between the $M$-Wright and the Mittag-Leffer functions \cite{MainardiF}:
 \begin{equation}
 M_\alpha(r)\div E_\alpha(-s)\,,\quad r \in R^+\,,
 \end{equation}
thus:
 \begin{equation}
  \int_0^\infty e^{-rt^\alpha}M_\alpha(r)dr=E_\alpha(-t^\alpha)\,,
 \end{equation}
 applying the change of variables $q=rt^\alpha$ we have:
  \begin{equation}
  \int_0^\infty e^{-q}\frac{1}{t^\alpha}M_\alpha(q/t^\alpha)dq=E_\alpha(-t^\alpha)\,.
 \end{equation}

Applying this idea to the most general solution for CTRW \cite{montroll1964,scalas_etal-pre-2004} is it possible to write it in terms of a superposition of Markovian
components, each characterized by the same jump PDF \cite{DiTullio-MT}:
\begin{equation}
 P(r,t)=\int_0^\infty P_M(r,t/\tau)f(\tau,t)d\tau=\int_0^\infty P_M(r,q)\mathcal{H}(q,t)dq\,.
\end{equation}

The simplest diffusion process of molecules associated to the transmembrane potential solution of the classic cable equation is $P_M'(r,t)=P_M(r,q)e^{-q}$,
with $P_M(r,q)=\frac{1}{\sqrt{4\pi q}}e^{-r^2/4q}$, standard~diffusion process, multiplied by the exponential factor $e^{-q}$ that accounts for the loss of particles in the system.
Following the same superposition principle after turning on the exponential decay term, the transmembrane potential $P(r,t)$ corresponds to the 
integral of the solution of the classic cable equation averaged 
by the same $\mathcal{H}(q,t)=\frac{1}{t^\alpha}M_\alpha(q/t^\alpha)$, by considering that the decay is subjected to the same timescale of the diffusion process:

\begin{equation}\label{intform}
  P(r,t)=\int_0^\infty \frac{1}{\sqrt{4\pi q}}e^{-(\frac{X^2}{4q}+q)}\frac{1}{t^\alpha}M_\alpha(\frac{q}{t^\alpha})dq\,.
\end{equation}

The classic problem was written in terms of the adimensional variable $T=t/\tau$, with $\tau=c_m r_m$, and $X=x/\lambda$, with $\lambda=\sqrt{r_i/r_m}$
related to the circuit component of the membrane element.
The~solution of the fractional process can be written in terms of a superposition of the classic solution weighted by the 
distribution of the circuit element parameters, thus we have:
\begin{equation}\label{solchauchy2}
 V_\alpha(x,t)=\int_0^\infty \frac{1}{\sqrt{4\pi t/\tau}}e^{-\left(\lambda^2\frac{x^2}{4t/\tau}+t/\tau\right)}
 \frac{t^{1-\alpha}}{\tau^2}M_\alpha(t^{1-\alpha}/\tau) d\tau\,.
 \end{equation}

 Then in terms of circuit elements the system results characterized by a capacitance that varies between the elements of the circuit according to a certain time dependent 
 distribution, considering $r_i, r_m$ unitary constant for simplicity, in the present case it corresponds to:
 \begin{equation}
  f(c_m,t)=\frac{t^{1-\alpha}}{c_m^2}M_\alpha(t^{1-\alpha}/c_m)\,.
\end{equation}

If there exists also a population of $r_m$, representing the transmembrane resistance, the time decay of the solution and diffusion processes 
are described by two different but correlated distributions, because~the coefficient $r_m$ disappears in the Gaussian factor.

In the comb-like model, the timescale is the average sojourn time in the spine and can be related to the geometry of the spine:~{a simple geometrical approximation involves spines composed by a~head of volume $V$ connected to the backbone by a~neck of
cylindrical shape with length $L$ and radius $a$. The mean time spent in such a spine is  $\tau = (LV )/(\pi a 2D)$,
where $D$ represents a quantity called }diffusivity of the spine \cite{Iomin-2013,Mendez-2013}. 
If this volume may change dynamically it makes sense to consider a~time-dependent distribution in this case as well.

Equation (\ref{solchauchy2}) can also be interpreted within the ggBm approach \cite{Mura}; rewriting the integral form as follows:
\begin{equation}
 V_{\alpha}(X,T)=\int_0^\infty \frac{1}{\sqrt{4\pi \Lambda T^\alpha}}e^{-\left(\frac{X^2}{4\Lambda T^\alpha}+\Lambda T^\alpha\right)}M_\alpha(\Lambda) d\Lambda\,,
\end{equation}
where inside the integral we recognize the fundamental solution expressed in Equation (\ref{fbm}) of the fBm model defined in Equation (\ref{cablefbm}) for the particular case $\alpha=\kappa$.

The ggBm-like stochastic process can be defined by the product:
\begin{equation}
 X'(t)=\sqrt{D}X(t)
\end{equation}
where $X(t)$ is a Gaussian process with unitary coefficient of diffusion, rescaled by the diffusion coefficient $D$ distributed according to:
\begin{equation}
 \rho(D,t)=\frac{1}{t^{\alpha-1}}M_\alpha(D/t^{\alpha-1})\,,
\end{equation}
where comes natural the change of variables $D=\Lambda t^{\alpha-1}$, which is the fBm definition of the diffusion coefficient, thus $p(\Lambda)=M_\alpha(\Lambda)$ \cite{Molina}.
The survival probability of each particle is conditioned to its diffusion coefficient $D$:
\begin{equation}
 \quad r(D,t)=e^{-D t}
\end{equation}

The partial differential equation (PDE) 
 for these processes can be derived by computing the Laplace-Fourier transform of the integral form in Equation (\ref{intform}), that reads
\begin{equation}
 \hat{\tilde{V_\alpha}}(s,k)=\frac{2s^{\alpha-1}}{s^\alpha+1+k^2}\,,
\end{equation}
thus the transformed PDE is 
\begin{equation}
 2s^{\alpha-1}=s^\alpha \hat{\tilde{V_\alpha}}(s,k)+ \hat{\tilde{V_\alpha}}(s,k)+k^2 \hat{\tilde{V_\alpha}}(s,k)\,,
\end{equation}
which correspond to the time fractional cable equation described in \cite{Vitali} with $0<\alpha<1$ for Cauchy initial conditions:
\begin{equation}\label{frac}
\frac{\partial^\alpha V_\alpha(X,T)}{\partial T^\alpha} =\frac{\partial^2 V_\alpha(X,T)}{\partial X^2} - V_\alpha(X,T) \,,
\end{equation}
where $\frac{\partial^\alpha}{\partial T^\alpha}$ is the Caputo time fractional derivative.

\section{Discussion}

Fractional calculus is often used to catch by parsimonious mathematical approach some underlying complex behavior.
Caputo's fractional derivative is a non-local operator and for this reason, as pointed out in \cite{Teka},
it could be introduced to explain emergent behaviors such as the appearance of multiple timescale dynamics  
and memory effects, related to the complexity of the medium.~{In this work we derived two possible stochastic processes, CTRW and ggBm, for inert tracer diffusion in spiny dendrites that in principle give rise to the same partial
differential equation for the transmembrane potential.
The physical mechanisms expected behind CTRW and ggBm are strictly related to the process construction.
Underlying the CTRW model there is the concept of trapping in the spines, where the distribution of timescales account the variability in the geometry and size of these spines.
Underlying the ggBm model there is the idea that the environment is dynamical and that each particle may feel the surround in a different way. 
Both the approaches are approximations of the real system, and the aim of these models is to describe data behavior and possibly predict interesting biological features.}

The PDF evolution of both the processes is described by the time fractional generalization of the cable equation presented in Equation (\ref{frac}), 
that can be solved for the most common boundary and initial
conditions by the application of the Efros theorem of Laplace transforms \cite{Vitali}. The transmembrane potential function 
described by this model fulfills many of the biological features that have been previously suggested to explain spines role, as the compensation of delay in postsynaptic potentials
and the attenuation time of the signal.

The first process is a CTRW built as a superposition of Markovian processes, each one subjected to a different timescale of the waiting 
time distribution, where the timescale follows a non stationary distribution.
This means that the whole system change in time modifying the profile of the timescale distribution.
The second process is based on a ggBm-like approach in which a Brownian process with unitary diffusion coefficient is rescaled 
by a random scale that is non stationary distributed as well. This scale represents in fact the value of the diffusion coefficient
and can be used to generate anomalous time-scaling of the MSD in the final variable:
\begin{equation}
 \langle X'^2\rangle = \langle D \rangle \times t\,.
\end{equation}

If $r(D,t)=1$ we obtain $\langle X'^2\rangle\sim t^\alpha$.
The exponential suppression has the effect that probability distribution collapse to zero in the infinite time, because all the particles disappear from the system. 

Except for the exponential suppression that accounts for loss of particles, a similar non-stationary ggBm process has been proposed in \cite{Molina} 
as an alternative to CTRW to account the Ergodicity Breaking (EB)
described by several experiments on diffusion of cellular components in living systems. Despite~both ggBm and CTRW may account for EB, in \cite{Molina} 
it was shown that the $p$-variation test 
provides different values for the two alternative processes, and that values obtained for ggBm where compatible with the experimental dataset considered 
in their research, in contrast to CTRW.

For these reasons it seems promising to characterize the present processes looking forward for single particle tracking data to be compared with the models.
Moreover, the two processes presented here
account for the complexity of the phenomena directly from geometrical (waiting time timescales distribution) and/or electrophysiology (cell resistances 
and capacitance values distributions) properties of the system,
that could be directly measured as it was done for spine density profiles in \cite{Santamaria-2011}.~Finally, the anomalous transport phenomena 
is generated by a proper superposition of classic processes, that is not ad-hoc
but can be related to experimental observations, 
clearly simplifying also the computational efforts of the simulation procedures of the trajectories. 

\vspace{6pt} 

\section*{Aknowledgement}
The work of 
Francesco Mainardi has been carried out in the framework of the activities of the
 National Group of Mathematical Physics (INdAM-GNFM).
All the authors acknowledge support by the Italian Ministry of Education and the 
Interdepartmental Center  "Luigi Galvani" for integrated studies of Bioinformatics,
Biophysics~and Biocomplexity of the University of Bologna. We also thanks for useful discussions G. Pagnini (BCAM, Spain).

Silvia Vitali, Francesco Mainardi and Gastone Castellani wrote the paper; Gastone~Castellani and Silvia Vitali accounts for the biological part, 
Silvia Vitali and Francesco Mainardi for the~analytics.

The authors declare no conflict of interest.

\section*{Abbreviations}

\noindent 
\begin{tabular}{@{}ll}
Bm & Brownian motion\\
CTRW & Continuous Time Random Walk\\
ggBm & generalized grey Brownian motion\\
fBm & fractional Brownian motion\\
NPE & Nernst Planck Equation\\
PDF & Probability Density Function\\
PDE & Partial Differential Equation\\
EB & Ergodicity Breaking
\end{tabular}


\end{document}